\documentclass[%
 reprint,
 amsmath,amssymb,
 aps,
]{revtex4-2}

\usepackage{graphicx}
\usepackage{dcolumn}
\usepackage{twistor}
\usepackage{package-notes}
\usepackage{bm}
\usepackage{simpler-wick}
\usetikzlibrary{tikzmark}
\usepackage{amsmath}
\usepackage{mathrsfs}
\usepackage{tikz-cd}
\usepackage{adjustbox}

\newcommand{\dbeta}{\Dot{\beta}}

\newcommand{\ba}{\boldsymbol{a}}
\newcommand{\bb}{\boldsymbol{b}}
\newcommand{\bc}{\boldsymbol{c}}

\begin{document}

\preprint{APS/123-QED}

\title{2d theory for asymptotic dynamics \\
of 4d (self-dual) Einstein gravity}

\author{Wei Bu}
 \affiliation{University of Edinburgh \& Northeastern University}
\author{Sean Seet}%
\affiliation{%
 DAMTP, University of Cambridge \& University of Edinburgh
}%

\date{\today}

\begin{abstract}
In this paper, we present a simple chiral 2d theory living on a momentum space celestial sphere whose behaviour exactly produces various IR dynamics of recent resurged interests for 4d (self-dual) Einstein gravity in asymptotically flat spacetimes. We demonstrate how to use simple 2d CFT computations to reproduce 4d BMS algebra and $w_{1+\infty}$ algebra, deduce the form of both chiral and anti-chiral stress tensors and recover the necessity for dressing hard particles asymptotically with soft modes. We further discuss how possible extensions of this 2d theory incorporates further dynamical information of 4d Einstein gravity. 
\end{abstract}

\maketitle

\section{Introduction}
Recent developments in the study of infrared (IR) physics in 4d asymptotically flat spacetimes has reignited the interests in further studying asymptotic degrees of freedom in such spaces. Both in gauge theory and gravity, large $r$ expansion of the field strength/metric has played a crucial role in understanding the residual asymptotic degrees of freedom in these theories. They are usually referred to as soft modes or radiation modes in the literature \cite{Weinberg:1965,Bondi:1962px,Sachs:1961zz,Newman:1966ub,Newman:1976gc}, we shall stick to this name throughout this paper. In particular in general relativity, this was done both from the perspectives of phase space and dynamical on-shell observables such as wavefunctions and scattering amplitudes \cite{Pasterski:2017kqt,Pasterski:2016qvg,Casali:2022fro,Banerjee:2020zlg,Adamo:2023zeh,Sahoo:2020ryf,Sahoo:2021ctw,Strominger:2017zoo}. The symmetry principles organizing these soft degrees of freedom are perhaps surprisingly the asymptotic BMS groups \cite{Bondi:1962px,Sachs:1961zz}, which has a closely related cousin acting on the level of special kinematical limits of scattering states, the $\mathcal{L}w_{1+\infty}$ algebra \cite{Strominger:2021mtt,Guevara:2021abz,Adamo:2021lrv}. By Mellin transforming massless on-shell data like scattering states and IR limit of scattering amplitudes, they transform under the 2d conformal group explicitly. An extensive part of the literature focused on studying such Mellin transformed amplitudes which to an extent, constraints the form of the dynamics of the asymptotic degrees of freedoms of 4d Einstein gravity. Unlike traditional conformal field theory (CFT)s, it is not clear how essential properties such as unitarity can be manifested just by observing these on-shell conformal objects. 

It is slightly dissatisfying that an explicit action principle is not present, which would dramatically enhance the understanding of these radiation degrees of freedom of 4d gravity. Although there has been attempts to write down field theories for simple theories such as abelian  \cite{He:2024ddb,Kapec:2022xjw} and non-abelian \cite{Magnea:2021fvy} gauge theories, from spacetime itself using usual AdS/CFT intuitions \cite{deBoer:2003vf,Cheung:2016iub,Kim:2023vbj} and attempts to reproduce individual helicity graded amplitudes \cite{Melton:2024akx}. It is rather difficult to infer and connect these with the aforementioned dynamics observed in theories in 4d asymptotically flat spacetimes. In this paper, we write down an explicit Lagrangian on a 2-sphere and demonstrate how to extract known IR phenomenon. The 2d action was originally derived using twistor space techniques starting from the self-dual sector of 4d Einstein gravity by the two authors \cite{Bu:2024cql}, we intend to omit the detail and refer interested readers there. 

In section \ref{sec:level1}, we write down the 2d chiral action which resembles the form of a principal chiral model and go into detail establishing notations and a map to 4d Minkowski space. In section \ref{sec:ST/SR}, as a sanity check, we first demonstrate how vectors generating the BMS group (super-translation and super-rotation) can be extracted from the 2d theory. In section \ref{sec:stress_tensor}, we elucidate how both holomorphic and anti-holomorphic part of the stress tensor appear in the 2d conformal theory, with their actions on a generic conformal primary. In section \ref{sec:dressing}, we use the action of a soft current (closely related to the superrotation current) to further deduce the necessity for dressing gravitationally coupled on-shell states with soft cloud, simply by requiring $U(1)$ gauge invariance of our action. 

\section{\label{sec:level1} 2d action and interpretation}
Consider a principal chiral model (PCM)-like action on a Riemann sphere with the (Jacobi) structure constants $f^{\boldsymbol{a}_1\boldsymbol{a}_2}_{\boldsymbol{a}_3}$ and non-degenerate metric $\kappa^{\boldsymbol{a}_1\boldsymbol{a}_2}$ normalized as $\kappa^{\boldsymbol{a}_1\boldsymbol{a}_2}\kappa^{\boldsymbol{c}\boldsymbol{d}}\kappa_{\boldsymbol{c}\boldsymbol{d}} = f^{\boldsymbol{a}_1\boldsymbol{e}}_{\boldsymbol{b}} f^{\boldsymbol{a}_2\boldsymbol{b}}_{\boldsymbol{e}}$ 
\begin{align}\label{2d_action}
        S := \int_{\mathbb{CP}^1} \d z \wedge\left(\Phi_{\boldsymbol{a}}\bar\partial \alpha^{\boldsymbol{a}} +\Lambda_{\boldsymbol{a}}\bar\partial \beta^{\boldsymbol{a}} + \right. \nonumber
        \\ \left. h^{\boldsymbol{a}} J_{\boldsymbol{a}} + \tilde h^{\boldsymbol{a}} \tilde J_{\boldsymbol{a}}\right)\,.
\end{align}
The index $\boldsymbol{a}$ and the structure constants are determined by the theory we intend to describe. The fields $\Phi_{\boldsymbol{a}}, \alpha^{\boldsymbol{a}}, \Lambda_{\boldsymbol{a}}, \beta^{\boldsymbol{a}}$ indexed by $\boldsymbol{a}$ are sections of $\mathcal{O}(n)$ bundles on $\mathbb{CP}^1$ with $\boldsymbol{a}$ dependent weights $\Omega^{0}(\mathbb{CP}^1, \mathcal{O}(*))$, while the fields $h^{\boldsymbol{a}}, \tilde h^{\boldsymbol{a}}$ are $(0,1)-$forms $\in \Omega^{0,1}(\mathbb{CP}^1, \mathcal{O}(*))$. The currents $J_{\boldsymbol{a}},\tilde J_{\boldsymbol{a}}$ are the following combinations of the fundamental fields $\Phi_{\boldsymbol{a}}, \alpha^{\boldsymbol{a}}, \Lambda_{\boldsymbol{a}}, \beta^{\boldsymbol{a}}$, and $c(\boldsymbol{a})$ is a function of the label $\boldsymbol{a}$:
\begin{equation}\label{eq: currents}
    \begin{aligned}
    J^{\boldsymbol{a}} := &c(\boldsymbol{a})\kappa^{\ba\bb}\Phi_{\bb} + f^{\ba\bb}_{\bc}\left(\Phi_{\boldsymbol{b}} \alpha^{\boldsymbol{c}}+ \Lambda_{\boldsymbol{b}}\beta^{\boldsymbol{c}}\right)\,,
     \\
     \tilde J^{\boldsymbol{a}} := &c(\boldsymbol{a})\kappa^{\boldsymbol{a}\boldsymbol{b}}\Lambda_{\boldsymbol{b}} - f^{\ba\bb}_{\bc}\Lambda_{\boldsymbol{b}} \alpha^{\boldsymbol{c}}\,.
    \end{aligned}
\end{equation}
In local coordinates $z$ on a local patch of $\mathbb{CP}^1$, the holomorphic free fields $\Phi_{\ba}$, $\alpha^{\ba}$, $\Lambda_{\ba}$ and $\beta^{\ba}$ obey the fundamental OPE relations:
\begin{equation}\label{ff_OPE}
    \Phi_{\boldsymbol{a}_1}(z_1)\alpha^{\boldsymbol{a}_2}(z_2)\sim \frac{\kappa_{\boldsymbol{a}_1}{}^{\boldsymbol{a}_2}}{z_1-z_2} \,,\quad    \Lambda_{\boldsymbol{a}_1}(z_1)\beta^{\boldsymbol{a}_2}(z_2)\sim \frac{\kappa_{\boldsymbol{a}_1}{}^{\boldsymbol{a}_2}}{z_1-z_2}
\end{equation}
with all other OPEs being regular. Using \eqref{ff_OPE}, Wick contractions between the bilinear currents give the OPE rules.
Setting $c(\boldsymbol{a})$ to be 0 for the moment, it can be seen that the currents obey the OPE rules 
\begin{align}
    &J^{\boldsymbol{a}_1}(z_1,\bar z_1) J^{\boldsymbol{a}_2}(z_2, \bar z_2) \sim \nonumber 
    \\
    &\frac{\kappa^2 \kappa^{\boldsymbol{a_1a_2}}}{(z_1-z_2)^2} + \frac{f^{\boldsymbol{a_1a_2}}_{{\boldsymbol{a}_3}} J^{\boldsymbol{a}_3}(z_2, \bar z_2)}{z_1-z_2} + \,\cO\left((z_{12})^0\right)\label{JJOPE}
    \\
    &J^{\boldsymbol{a}_1}(z_1,\bar z_1) \tilde J^{\boldsymbol{a}_2}(z_2, \bar z_2) \sim \frac{f^{\boldsymbol{a_1a_2}}_{{\boldsymbol{a}_3}} \tilde J^{\boldsymbol{a}_3}(z_2, \bar z_2)}{z_1-z_2}+\,\cO\left((z_{12})^0\right)
    \\
    &\tilde J^{\boldsymbol{a}_1}(z_1,\bar z_1) \tilde J^{\boldsymbol{a}_2}(z_2, \bar z_2) \sim  \,\cO\left((z_{12})^0\right)\,.
\end{align}
With specific choices of the domain of $\boldsymbol{a}$, and the data $c(\boldsymbol{a}),\kappa^{\boldsymbol{a}_1\boldsymbol{a}_2},f^{\boldsymbol{a}_1\boldsymbol{a}_2}_{\boldsymbol{a}_3}$, it was shown in \cite{Bu:2023cef, Bu:2024cql} that the $J, \tilde J$ OPEs are as given above (this is a nontrivial constraint on $c(\boldsymbol{a})$), giving the asymptotic symmetry algebra of Yang-Mills, the \textit{S-algebra} \cite{Bu:2023cef} or the asymptotic symmetry algebra of gravity, (including the \textit{w-algebra}) \cite{Bu:2024cql} from this PCM-like action. 

The case of interest for this paper is a sector of self-dual gravity, as derived and studied in \cite{Bu:2024cql}. A sketch of the derivation is presented in Appendix \ref{appendix: twistor derivation}. The index $\boldsymbol{a}:=(\Delta,s,k)$ appearing in an upstairs position takes values in $(1+ \im \mathbb{R},\mathbb{Z}, \mathbb{Z})$, while the metric and structure constants are defined as
\begin{align}\label{eq: algebra constants}
    &\kappa^{\boldsymbol{a}_1\boldsymbol{a}_2} := \delta(\Delta_1+\Delta_2)\delta_{s_1+s_2,0} \delta_{k_1+k_2,0} 
    \\
    & f^{\boldsymbol{a}_1\boldsymbol{a}_2\boldsymbol{a}_3} := \delta(\Delta_1+\Delta_2+\Delta_3-1)\delta_{(s_1+s_2+s_3-1),0}\times \nonumber
    \\ &\delta_{(k_1+k_2+k_3-1),0}\left(k_2\left(\frac{s_3+\Delta_3}{2}\right) - k_3\left(\frac{s_2+\Delta_2}{2}\right)\right)
    \\ & c(\boldsymbol{a}):= \frac{-\Delta-s}{2}\,.
\end{align}
Index contraction is an integral over $\Delta$ and a sum over $s, k$. The choice of signs in the metric $\kappa^{\boldsymbol{a}_1\boldsymbol{a}_2}$ means that downstairs $\boldsymbol{a}$ indices take values in $(1+\im \mathbb{R},\mathbb{Z}, \mathbb{Z})$. The fields $\Phi,\Lambda$ are required to have $k\in\mathbb{Z}^+$, while the other fields $\alpha^{\ba}$, $\beta^{\ba}$, $h^{\ba}$ and $\tilde h^{\ba}$ can have arbitrary values of $k$. Note however, the fields $\alpha^{\ba}$, $\beta^{\ba}$, $h^{\ba}$ and $\tilde h^{\ba}$ with positive $k$ are non-dynamical and have no nontrivial OPEs. This is because they do not enter into the kinetic term. 
After defining the structure constant, we recognize that \eqref{JJOPE} contains as a closed subsector the $w_{1+\infty}$ algebra discovered in the asymptotic symmetry group of 4d self-dual gravity.

The factor of $\kappa^2:= \kappa^{\boldsymbol{a_1a_2}} \kappa_{\boldsymbol{a_1a_2}}\propto \text{vol}(M)$ in the double pole of the $JJ$ OPE therefore diverges as the volume of the Minkowski space $\text{vol}(M)$. Physically, this is because the indices $\boldsymbol{a}$ label (infinitely many) mode numbers in a mode decomposition of 4d spacetime fields. This divergence needs to be regulated, for example with a cutoff (i.e restricting the allowed mode numbers), as was done in \cite{Bu:2023cef,Bu:2024cql}.

Since the $J$ current is gauged, the double pole in the $JJ$ OPE indicates the quantum mechanical failure of gauge-invariance. This can be removed by a simple 2d "Green-Schwarz" type mechanism to cancel the anomaly and recover a quantum mechanically consistent theory. One simply adds a non-chiral bosonized term to the action
\begin{equation}
    \int_{\mathbb{CP}^1}\,\partial\zeta^{\ba}\bar\partial\zeta_{\ba}+\im h^{\ba}\partial\zeta_{\ba}\,.
\end{equation}
This alters the current coupling to $h$ by a term whose single contraction cancels the two-point bubble diagram. It is curious what the interpretation of the $\zeta_{\ba}$ is on regular twistor space and how it relates to the usual chiral anomaly \cite{Bittleston:2022nfr,Costello:2021bah}.


Physically, the PCM-like action is best understood as a free-field realization \footnote{Although free-field realizations of these asymptotic symmetry algebras can be conjured with no canonical choice, the PCM-like actions in \cite{Bu:2023cef,Bu:2024cql} are distinguished by being directly derived from twistor actions which have been proven to be classically equivalent to spacetime theories of self dual Yang-Mills, full Yang-Mills, and self dual gravity.} of the asymptotic symmetry algebra of self-dual gravity (resp. of self-dual Yang-Mills in \cite{Bu:2023cef}). The symmetry algebra described by \eqref{JJOPE} includes (by restricting mode numbers to $Im(\Delta) = 0$) but is much larger than the expected loop algebra of $w_{1+\infty}$. This is because the symmetry algebra in \eqref{JJOPE} is unbroken, as we have made no choices of gauge or of boundary/falloff conditions. Choices of falloff conditions, regularity assumptions, and partial gauge fixings would break the symmetry algebra in \eqref{JJOPE} down to smaller subalgebras. These would present as constraints on the allowed mode numbers.

To make contact with self-dual gravity as perturbations around real Lorentzian spacetime, the $\mathbb{CP}^1$ the action \eqref{2d_action} is defined on is best understood as the projectivisation of the nonzero real Lorentzian null momenta, which is called the momentum-space celestial sphere:
\begin{equation}
    \mathbb{CP}^1_{\lambda} \cong S^2 = \frac{\{k^2 =0 | k^\mu \in \mathbb{R}^{3,1}\}\setminus \{k^{\mu}=0\}}{k^\mu \sim r k^\mu, \quad r \in \mathbb{R}^*}\,.
\end{equation}  
To relate to Minkowski space with the origin removed $\mathcal{M}^{\cO}:=\{\mathcal{M}\setminus\{x^{\mu}=0\}\}$\footnote{We make an explicit choice of the origin and remove it from Minkowski space.}, we ask the mode numbers $\boldsymbol{a}=(\Delta,s,k)$ to label certain behaviors of fields on $\mathcal{M}^{\cO}$ as we shall elaborate on now.
The 4 components of $x^{\mu}$ can be rewritten as a matrix by multiplying the Pauli matrices $x^{\mu}=\sigma^{\mu}_{\alpha\dal}x^{\alpha \dot \alpha} \in \mathcal{M}^{\cO}$, which can be separated into a complex conjugate pair and two real coordinates by projecting them onto the basis defined by coordinates of $\mathbb{CP}^1_{\lambda}$\footnote{In different signatures, the conjugation relations between the four projected components are different, in Lorentzian signature, $q$ and $\bar q$ are complex conjugates while $v$ and $v'$ are real and independent.}:
\begin{equation}\label{basis}
    \{x^{\alpha \dot \alpha}\} = \{\underbrace{([\hat{\bar \lambda} | x |\lambda \ra, [\bar \lambda | x | \hat \lambda \ra)_{\mathbb{C}^*}}_{q,\bar q}, \underbrace{[\bar \lambda |x |\lambda \ra}_{v}, \underbrace{[\hat{\bar\lambda}| x| \hat\lambda\ra}_{v'}\}\,,
\end{equation}
where in components, the 2-spinors are defined as $\lambda^{\alpha}=\binom{\lambda_0}{\lambda_1}$, $\bar\lambda^{\dal}= \binom{\bar\lambda_0}{\bar\lambda_1}$, $\hat{\lambda}^{\alpha}=\binom{-\bar\lambda_1}{\bar\lambda_0}$ and $\hat{\bar\lambda}^{\dal}=\binom{-\lambda_1}{\lambda_0}$. The spinor helicity notation with mixed angle and square brackets is denoting $\la\zeta|x|\bar\zeta]=\zeta^{\alpha}\,x_{\alpha\dal}\bar\zeta^{\dal}$. $(\Delta,s)$ are mode numbers for a $\mathbb{R}^+ \times U(1) = \mathbb{C}^*$ action on $\mathcal{M}^{\cO}$, with orthogonal basis functions $B_{\Delta,s}$:
\begin{equation}
    B_{\Delta,s}:=(q\bar q)^{-\Delta/2} \left(\frac{\bar q}{q}\right)^{s/2} = q^{-\left(\frac{\Delta+s}{2}\right)}\bar q^{-\left(\frac{\Delta-s}{2}\right)}\,.
\end{equation}
Orthogonality is defined with respect to the natural inner product
\begin{equation}
\begin{aligned}
    &\int_{\mathbb{C}^*} \d q \d \bar q\, B_{\Delta_1,s_1}B_{\Delta_2,s_2} 
    \\
    =& \int_{\mathbb{R}^+ \times U(1)} r \d r \d \theta \,r^{-\Delta_1-\Delta_2} e^{\im \theta(-s_1-s_2)}
    \\
    =&2 \pi \delta_{s_1+s_2,0}\int_\mathbb{R} \d \rho e^{(2-\Delta_1-\Delta_2)\rho}, \quad \rho = \log r \,.
\end{aligned}
\end{equation}
$\Delta$, the mode number for the $\mathbb{R}^+$ dilatation, is required to take values in $1+\im\mathbb{R}$ for normalizability with respect to the integral over $\mathbb{R}^+$ \cite{Bu:2024cql}. $s$, the Fourier mode number for the $U(1)$, is required to take values in $\mathbb{Z}$ for single-valuedness. 

On the other hand, $k$ is not quite a mode number, but rather a positive integer encoding the strength of the singularity of the free fields $\Phi, \Lambda$ that have singular behavior $1/v^k$ at $v=0$. Physically, they represent large diffeomorphism degrees of freedom for the gravitons. Being large diffeomorphisms, they signal the presence of arbitrary boundary value the bulk gravitons can asymptote to with generic singular behavior along $\mathscr{I}$. Rather than using mode functions orthogonal with respect to an inner product, the orthogonality rule for $k$ arises from the canonical pairing of functions and distributions
\begin{equation}
    \int \d v v^{k_1-1} \delta^{(k_2-1)}(v)
    = (-1)^{k_2-1} (k_2-1)! \delta_{k_1,k_2}\,.
\end{equation}
The Minkowski space dependence is encoded in the triplet of mode numbers $\ba=(\Delta,s,k)$ labeling each field on $\mathbb{CP}^1$. The reason that integrating out the 4d spacetime factor $\mathcal{M}^{\cO}$ gives 3 mode numbers rather than 4 is because the form of the fields $\Phi_k(x, \lambda), \Lambda_k(x, \lambda)$ are not general functions of $\{\mathcal{M}\setminus\{x^{\alpha \dot \alpha}=0\}\}$, but are independent of $v' := [\hat{\bar\lambda}| x| \hat\lambda\ra$ and $\delta$-function supported on $v = 0$. The reader is invited to refer to the paper \cite{Bu:2024cql} for more details, including the full first principles derivation of the action functional from the twistor space action for 4d self-dual Einstein gravity.


Following through the derivation, $\Phi_{\boldsymbol{a}}, \Lambda_{\boldsymbol{a}}$ are modes of the singular gauge transformations of the negative and positive helicity graviton respectively. $h^{\boldsymbol{a}}, \alpha^{\boldsymbol{a}}$ are modes of the positive helicity graviton, while $\tilde h^{\boldsymbol{a}}, \beta^{\boldsymbol{a}}$ are modes of the negative helicity graviton.

To make contact with more conventional descriptions, in the original theory on $\mathcal{M}^{\cO}\times \mathbb{CP}^1_\lambda$, $h^{\Delta,s,k}(\lambda)$ are modes of the graviton expanded in the basis $B_{\Delta,s}$ and $v^k$. Collectively $h(x, \lambda)= \int\d\Delta\sum_{s,k} h^{\Delta,s,k}B_{\Delta,s}\,v^k$ is a $(0,1)$-form with homogeneity $+2$ on $\mathbb{CP}^1$, which is related to a linearised positive helicity graviton on the spacetime factor by the integral transform over the celestial sphere \cite{Sparling1990,Penrose_Rindler_1988}:
\begin{equation}\label{Sparling}
    g_{\alpha\dal\beta\dbeta} = \int_{\mathbb{CP}^1_{u=0}}\D\lambda\wedge \frac{\hat\lambda_{\alpha}\hat\lambda_{\beta}\hat{\lambda}_{\gamma}\hat{\lambda}_{\rho}}{\la \lambda\hat{\lambda}\ra^4}\,\frac{\partial^2}{\partial x_{\gamma}{}^{\dal}\partial x_{\rho}{}^{\dbeta}}h(x,\lambda)\,.
\end{equation}
One would like to stress that $h$ and $\tilde h$ are defined up to local gauge transformations:
\begin{equation}\label{gauge_trans}
    h\to h+ \bar\partial \chi + \left\{h,\chi\right\} = h +\bar\D\chi\,,
\end{equation}
with the Poisson bracket written in $x^{\alpha\dal}, \lambda^\alpha$ coordinates as
\begin{equation}
    \{\cdot\,,\,\cdot\} = \frac{\hat \lambda^{\alpha}\hat \lambda^\beta}{\la \lambda \hat \lambda \ra^2}\,\frac{\partial\,\cdot}{\partial x^{\alpha\dal}}\,\frac{\partial\,\cdot}{\partial x^{\beta}_{\dal}}\,.
\end{equation}
We shall further clarify these local gauge transformations include the asymptotic diffeomorphisms of Minkowski space. 

As a final comment on the space $\cM^{\cO}\times\mathbb{CP}^1_{\lambda}$, it is important to note that the momentum space celestial sphere $\mathbb{CP}^1_{\lambda}$ is distinct from the position space celestial sphere, described in Bondi coordinates by $\mathbb{CP}^1_{w,\bar w}$ in the coordinatisation of Minkowski space by $\mathbb{R}^{+}_{r}\times\mathbb{R}_{u}\times\mathbb{CP}^1_{w,\bar w}$ \cite{Bondi:1962px,Sachs:1961zz}:
\begin{equation}
    x^{\alpha\dal} = T^{\alpha\dal} u+ r\,\frac{w^{\alpha}\bar w^{\dal}}{\la w\hat{w}\ra}\,,
\end{equation}
where $T^{\alpha\dal}$ is $\text{diag}(1,1)$. Note that the Bondi decomposition is different from the basis labels we have used to write down our 2d theory \eqref{basis}, the Bondi decomposition comes handy when one would like to express the results in more recognizable forms in the 
literature.

\subsection{The geometric correspondence}
We would like to further elucidate the relation between the 2d theory \eqref{2d_action} with infinitely many fields living on the momentum space celestial sphere $\mathbb{CP}^1_{\lambda}$ and its correspondence with self-dual gravity on Minkowski space $\cM^{\cO}$. Especially in terms of the off-shell fields we have written in the action functional. This goes through a correspondence space, which is simply the direct product of the two spaces: $\cM^{\cO}\times\mathbb{CP}^1_{\lambda}$. On $\cM^{\cO}\times\mathbb{CP}^1$, spacetime diffeomorphisms are encoded as cohomological gauge transformations, as defined via the integral transform \eqref{Sparling}. For self-dual gravity on $\cM^{\cO}$, the Wick rotation of the Mason-Wolf twistor action \cite{Mason:2005zm} gives a holomorphic QFT on the product space $\cM^{\cO}\times\mathbb{CP}^1_{\lambda}$:
\begin{equation}\label{twistor sdg}
    \int_{\cM^{\cO}\times\mathbb{CP}^1}\lambda^{\alpha}\lambda^{\beta}\d^2x_{\alpha\beta}\wedge\D\lambda\wedge \tilde h\wedge\left(\d h+\frac{1}{2}\{h,h\}\right)\,.
\end{equation}
This theory was shown to be classically equivalent to the spacetime self-dual gravity action in the case of $\mathbb{R}^4 \times \mathbb{CP}^1$, and the argument for $\mathcal{M}^{\cO} \times \mathbb{CP}^1$ is precisely identical. The relevant components of $\d$ that contribute to the kinetic term in this action are $\bar\partial_{\lambda}+\bar\partial_{q}+\d_{v} $.

\begin{figure}[!]
    \centering
    \begin{equation*}
\adjustbox{scale=1.43}{
\begin{tikzcd}[ampersand replacement=\&, column sep=0.02em]
        \&\& \cM^{\cO}\times\mathbb{CP}^1_{\lambda} \arrow[bend right=20, swap]{lldd}[sloped,above] {\text{\tiny large $r$}} \arrow[bend left=20, swap]{ddrr}[sloped,above] {\text{\tiny mode}}\\\\
        \cM^{\cO} \arrow[bend left=45,swap]{rruu}[sloped,above] {\text{twistor}} \&\&\&\& \mathbb{CP}^1_{\lambda} \arrow[bend right=45, swap]{uull}[sloped,above] {\text{recover basis}} \arrow[bend right=150,swap]{llll}[below,pos=.5] {\text{transform}}
\end{tikzcd}
}
\end{equation*}
    \caption{}
    \label{fig:1}
\end{figure}
In the reduction to the $\mathcal{M}^{\cO}$ self-dual gravity action, an intermediate step is integrating out the extraneous $\mathbb{CP}^1_{\lambda}$ coordinates $(\lambda^{\alpha},\bar\lambda^{\dal})$. At the level of the off-shell quantum fields, at the linearised level this can be shown to be equivalent to the integral transform \eqref{Sparling}. For the on-shell states relevant to Celestial holography, in the large $r$ limit, the position space Celestial sphere $\mathbb{CP}^1_{w}$ and the momentum space celestial sphere $\mathbb{CP}^1_{\lambda}$ are identified. In order to see this, we examine the following scattering eigenstate (i.e. hard particle), in Bondi coordinates:
\begin{equation}
    h_{\text{on-shell}}|_{L_x}=\frac{\Gamma(\Delta-1)\,\bar\delta(\lambda,\lambda_{p})\frac{\la\iota\lambda\ra^3}{\la\iota\lambda_{p}\ra^3}}{\underbrace{\left(u+r\la\lambda w\ra[\bar\lambda\bar w]/(\la w\hat{w}\ra\la\lambda\hat{\lambda}\ra)\right)^{\Delta}}}_{\left(\frac{\la\lambda|x|\bar\lambda]}{\la\lambda\hat{\lambda}\ra}\right)^{\Delta}}\,,
\end{equation}
where $\lambda_p^{\alpha}$ is the undotted spinor associated with the external momentum that the scattering eigenstate is associated with and $\iota^{\alpha}$ is an arbitrary reference spinor. We have suppressed the $\pm\im \epsilon$ in the denominator as it will play no role.
\begin{equation}
    p^{\alpha \dot \alpha} = \pm \omega \frac{\lambda_p^\alpha \bar{\lambda}_{p}^{\dal}}{\la \lambda_p \hat \lambda_p \ra}\,.
\end{equation}
$h_{\text{on-shell}}$ is the conformal primary eigenstate considered in the literature, which is the Mellin transform over energy scale of the usual momentum eigenstate. In the large $r$ limit, we can read off that the expression for $h_{\text{on-shell}}$ is suppressed by inverse powers of $r$, except in the case that $\la \lambda w \ra [\bar \lambda \bar w]$ vanishes. Specifically, for $Re(\Delta) > 1/2$, the $(u,r,w^\alpha)$ dependence takes the form
\begin{multline}
    \text{lim}_{r\to\infty}\,\frac{1}{\left(u+r\frac{\la\lambda w\ra[\bar\lambda\bar w]}{(\la w\hat{w}\ra\la\lambda\hat{\lambda}\ra)}\right)^{\Delta}}=
    \\ \sqrt{\frac{u}{r}}\frac{1}{u^{\Delta}}\delta(\rho)\frac{\sqrt{\pi}\Gamma(\Delta-1/2)}{\Gamma(\Delta)}+ \mathcal{O}(1/r)\,,
\end{multline}
where $\rho^2 = \frac{\la\lambda w\ra[\bar\lambda\bar w]}{\la w\hat{w}\ra\la\lambda\hat{\lambda}\ra}$. The delta function enforces the identification between the momentum space celestial sphere $\mathbb{CP}^1_{\lambda}$ and the position space celestial sphere $\mathbb{CP}^1_{w}$. Unlike the stationary-phase argument with plane waves, the large $r$-limit is defined and the identification of $\lambda^\alpha \sim w^\alpha$ does not require averaging over a thin shell of $(r, r+ \delta r)$ at large $r$.

Intuitively, this delta function appears for positive $\Delta$ in the large $r$ limit, as in the usual AdS/CFT scenario \cite{Witten:1998qj}. The reason the next order term is $\mathcal{O}(1/r^{3/2})$ is that roughly speaking, in the large $r$ limit, the expansion of the pole in powers of $r$ treats $\rho^2 r$ as an $\mathcal{O}(1)$ object (see Appendix \ref{appendix: large r delta functions} for more details).
At large $r$ therefore, the integral transform \eqref{Sparling} of $h_{\text{on-shell}}$ to a linearised metric perturbation on $\cM^{\cO}$ simplifies, as the integrand is localized to $\la \lambda w \ra = 0$.
Physically, the use of these on-shell scattering states and taking large $r$ has identified the momentum space celestial sphere coordinate $\lambda^\alpha$ with the Bondi coordinate $w^\alpha$.

In particular, readers familiar with the traditional asymptotic radiation analysis by Newman \cite{Newman:1976gc} might find this discussion illuminating, where the notion of Heavenly spaces is invented to recover deep interior metric from part of the complexified radiative data. The connection between our approach and the Heavenly spaces can be made precise through the asymptotic expansion of an on-shell wavefunction. Taking a linear superposition of $h_{\text{on-shell}}$ controlled by free profile functions $g_{\Delta} \in \Omega^{0,1}(\mathbb{CP}^1, \mathcal{O}(2))$
\begin{multline}
    h(u,r,w^\alpha, \lambda^\alpha) = \int_{1+\im \mathbb{R}} \d \Delta \frac{g_{\Delta}(\lambda, \bar \lambda)\la \lambda \hat \lambda \ra^\Delta}{\la \lambda x \bar \lambda]^{\Delta}}\xrightarrow{\text{large }r}
    \\
     \frac{1}{r} \partial_u^{-1}\left(\int_{1+\im \mathbb{R}} \d \Delta g_{\Delta}(w, \bar w)u^{-\Delta}\right)_{\lambda \sim w} + \cO(1/r^{3/2}) \,.
\end{multline}
Where the large $r$ behaviour is as a distribution integrated over a $\mathbb{C}$ patch of $\mathbb{CP}^1_\lambda$ (see Appendix \ref{appendix: large r delta functions} for the precise statement). We see that the effect of the large $r$ limit is to at leading order replace every appearance of
\begin{equation}
    \lambda^0/\lambda^1 \rightarrow w^0/w^1, \quad \frac{\la \lambda x \bar \lambda]}{\la \lambda \hat \lambda \ra} \rightarrow u
\end{equation}
and our on-shell fields are characterized by free data that depends only on $u$ and the $\mathbb{CP}^1$ coordinate $w^\alpha \sim \lambda^\alpha$. The partial complexification of radiative data Newman required comes from the signature of the bulk asymptotically flat space. For Euclidean signature, $\la \lambda x \bar \lambda] =: u_{\text{Newman}}$ is complex, hence the requirement for partial complexified $\scri_{\mathbb{C}}\cong\mathbb{C}_u\times\mathbb{CP}^1$. In our case, Lorentzian signature gives real $u$, which allows us to land on $\scri$ itself~\footnote{Strictly speaking there is an $\im \epsilon$ in the poles inherited from the original plane-wave scattering states that follows us through the Mellin transform, so that we have $g(w,\bar w)(u \pm \im\epsilon)^{\Delta}$ telling us which of $\scri^\pm$ we work on, although we will always suppress $\pm$ as it is largely irrelevant for our analysis of the gauge symmetries of self-dual gravity}.


Finally, as shown in the diagram figure \ref{fig:1}, if one would like to recognize 4d objects that correspond to off-shell objects on the 2d side, one would start with a 2d generically off-shell field $\Psi^{\ba}$ and pair it with its basis 
\begin{equation}
    \Psi(\lambda,x) = \int\d\Delta\sum_{s,k} \Psi^{\Delta,s,k}\,B_{\Delta,s}\,\frac{1}{v^k}\,.
\end{equation}
Then integrating out $\lambda$ in a prescribed way gives a space time field $\tilde\Psi(x)$.

\section{Supertranslations and superrotations}\label{sec:ST/SR}
In this section, we demonstrate how generators of the 4d BMS asymptotic symmetries can be extracted from local gauge transformations of the graviton representative. 

In this section, we demonstrate how generators of the 4d BMS algebra \cite{Bondi:1962px,Sachs:1961zz} can be extracted from currents and the OPE relations that they obey. In particular, supertranslation and superrotation are generated by the following currents:
\begin{equation}\label{eq: supertranslation and superrotation currents}
\begin{aligned}
    &f_{\text{ST}}:= 2\bar \partial_0 J^{0,0,1}-2 J^{1,1,0} \,,\\
    & Y_{\text{SR}} := -\frac{1}{2} \bar \partial^2_0 J^{0,0,-2}+ \bar \partial_0  J^{1,1,-1} - J^{2,2,0}\,,
\end{aligned}
\end{equation}
and the fusion rules of the charges they construct,
where notice that we have explicitly specified the three mode numbers $\ba=(\Delta,s,k)$ of the currents we are interested in. $\Delta$ outside the normalizability range $1+\im \mathbb{R}$ is expected for these currents, as the gauge transformations they correspond to are not normalizable. The derivatives are defined on the momentum space celestial sphere:
\begin{equation}
    \bar \partial_0 := \la \lambda \hat \lambda \ra \left\la \lambda \frac{\partial}{\partial \hat\lambda}\right\ra\,.
\end{equation}
In local coordinates $\bar\partial_0 = (1+z\bar z)z\frac{\partial}{\partial\bar z}$.

To understand the claim that the currents given above generate supertranslation and superrotations, we recall that the Ward identities relate current insertions with infinitesimal gauge transformations of the remaining operators.
The gauge transformations of the sum of all the graviton modes $h$ that generate the currents $f_{ST}, Y_{SR}$ are (recall that $q := \la \lambda x \hat{\bar\lambda}], v:=\la \lambda x \bar \lambda]/\la \lambda \hat \lambda \ra$)
\begin{equation}\label{eq: gauge parameters for ST, SR}
\begin{aligned}
    &h(\lambda, \bar \lambda) \xrightarrow{\delta_{\chi}} h(\lambda, \bar \lambda) + \epsilon \bar \D \chi
    \\
    &\delta h_{ST} = \bar \D \chi_{ST}:= \bar \D (-2(q+v \bar \partial_0) f(\lambda, \bar \lambda))
    \\ &\delta h_{SR} = \bar \D \chi_{SR}:= \bar \D \left(-\left(q^2 + q v \bar \partial_0 +\frac{1}{2} v^2 \bar \partial_0^2 \right)Y(\lambda, \bar \lambda)\right)\,,
\end{aligned}
\end{equation}
where we just recovered the modes indicated by the mode numbers on the current labels and $\bar\D = \bar \partial + \{h,\cdot\}$ is the covariant partial derivative defined in the gauge transformation \eqref{gauge_trans}. By the Ward identity argument above, the variation $\delta h_{ST}$ generates an insertion of $\int f_{ST} \wedge \bar \D \chi_{ST}$ in the action (and therefore at first order in $\epsilon$, an insertion of $\int f_{ST} \wedge \bar \D \chi_{ST}$ in a correlator), while $\delta h_{SR}$ in the action generates an insertion of $\int Y_{SR} \wedge \bar \D \chi_{SR}$. On spacetime, they correspond to infinitesimal diffeomorphisms
\begin{equation}
    h \xrightarrow{\delta_{\epsilon\chi}} h+ \epsilon \bar \D \chi
    \implies g_{\mu \nu} \rightarrow g_{\mu \nu} + \epsilon 2 \partial_{(\mu} V_{\nu)}\,,
\end{equation}
In the cases we shall consider in which $\bar \partial \chi \propto \D \hat \lambda$, the integral transform between $\chi$ and the vector field $V$ can be chosen to take the simple form
\begin{equation}\label{eq: vector field}
    V=V_{\alpha\dal}\partial^{\alpha\dal}=\int_{\mathbb{CP}^1}\frac{\D\lambda\D\bar\lambda}{\la\lambda\hat{\lambda}\ra^2}\,\underbrace{\frac{\hat{\lambda}_{\alpha}\hat{\lambda}^{\beta}}{\la\lambda\hat{\lambda}\ra^2}\,\frac{\partial}{\partial x^{\dal}{}_{\beta}}\chi(\lambda,x)\,\partial^{\alpha\dal}}_{\{\chi,\,\cdot\}} 
\end{equation}
see Appendix \ref{appendix: Penrose transform for vectors}.  Despite the suggestive form, it would be a mistake to conclude that commutators of the spacetime vector fields $V$ encode a Poisson algebra. This is because each of the Poisson brackets depend on the variables $\lambda^\alpha$ which are integrated over. Rather, commuted infinitesimal gauge transformations can be checked to generate the Poisson algebra
\begin{equation}
    h \xrightarrow{\delta_{\chi_1}\delta_{\chi_2}-\delta_{\chi_2}\delta_{\chi_1}} h + \epsilon_1\epsilon_2\bar \D\{\chi_1, \chi_2\} + \mathcal{O}(\epsilon^3)\,.
\end{equation}
For example, infinitesimal commuted superrotations generate the expected commutators of antiholomorphic vector fields $Y\bar \partial_0$
\begin{equation}
    [\delta_{\chi_{SR}(Y_1)},\delta_{\chi_{SR}(Y_2)}] = \delta_{\chi_{SR}(Y_1 \bar \partial_0 Y_2 - Y_2 \bar \partial_0 Y_1)} \,.
\end{equation}
By reading off their effect on on-shell fields, the $\chi_{ST},\chi_{SR}$ gauge parameters can be seen to be the gauge parameters that generate supertranslation and superrotation. It is useful to work with Bondi coordinates for the spacetime, in which the Poisson bracket should be written as
\begin{equation}\label{PB_x}
\begin{aligned}
    &\{\cdot\,,\,\cdot\} = -\frac{1}{2}\frac{\partial \cdot}{\partial q}\wedge\frac{\partial \cdot}{\partial v} =  \frac{\hat{\lambda}^{\alpha}\hat{\lambda}^\beta}{\la\lambda\hat{\lambda}\ra^2}\,\frac{\partial\,\cdot}{\partial x^{\alpha\dal}}\,\frac{\partial\,\cdot}{\partial x^{\beta}_{\dal}} \\
    &=\frac{1}{\la w\hat{w}\ra\la \lambda \hat \lambda \ra^2}\left(\frac{1}{r}\la\hat{\lambda} w\ra\left\la\hat{w}\frac{\partial}{\partial w}\right\ra-\la\hat{\lambda}\hat{w}\ra\frac{\partial}{\partial r}\right)\wedge\\
    &\left(-\frac{1}{r}\la\hat{\lambda}\hat{w}\ra\left\la w\frac{\partial}{\partial\hat{w}}\right\ra+\la\hat{\lambda} w\ra\left(2\frac{\partial}{\partial u}-\frac{\partial}{\partial r}\right)\right)\,.
\end{aligned}
\end{equation}Consider the action of the Poisson bracket on $h(\lambda,\bar \lambda,x)$, a generic on-shell state defined by free asymptotic data $g_{\Delta}(\lambda,\bar \lambda)\in \Omega^{0,1}(\mathbb{CP}^1,\mathcal{O}(2))$ and built up of the conformal primary wavefunctions
\begin{equation}
    h := \int_{1+\im \mathbb{R}} \d \Delta \,g_{\Delta}(\lambda,\bar \lambda)\left(\frac{\la \lambda \hat \lambda \ra}{\la\lambda|x|\bar\lambda]} \right)^{\Delta}\,.
\end{equation}
We have seen that the large $r$ limit of $1/[\bar \lambda x \lambda \ra^{\Delta}, Re(\Delta) >1/2 $ gives small $\la \lambda w \ra^2 \in \mathcal{O}(1/r)$. The leading order term in the expansion of the Poisson bracket acting on $h$ is therefore
\begin{multline}
    \{\cdot\,,\,h\} =\frac{1}{\la w\hat{w}\ra\la \lambda \hat \lambda \ra^2}\left(\frac{1}{r}\la\hat{\lambda} w\ra\left\la\hat{w}\frac{\partial}{\partial w}\right\ra \cdot \right)
    \left(2\la\hat{\lambda} w\ra\frac{\partial}{\partial u}h\right) \\- (\cdot \leftrightarrow h) + \mathcal{O}\left(\frac{1}{r^{3/2}}\right)\,.
\end{multline}
Substituting the supertranslation gauge parameter $\chi_{ST}$ into the Poisson bracket and taking the leading order term, we have the homogenous transformation of $h_{\Delta}$ under $\chi$ as
\begin{multline}
    \{\chi_{ST},h\} = \left\{r\frac{\la\lambda w\ra[\hat{\bar\lambda}\bar w]}{\la w\hat{w}\ra} f(\lambda, \bar \lambda), h \right\} \\
    = \frac{\la \hat \lambda w\ra^2\la \lambda \hat w \ra [\hat{\bar \lambda}\bar w]}{\la w \hat w \ra^2\la \lambda \hat \lambda \ra^2}f(\lambda, \bar \lambda)\left(\frac{\partial}{\partial u}\right)h +  \mathcal{O}\left(\frac{h_{\Delta}}{r^{1/2}}\right)\,.
\end{multline}
Importantly the on-shell states $h_{\Delta}$ do not depend on $q$. In the large $r$ limit, $\lambda^\alpha \rightarrow w^\alpha$ and the prefactor is in fact independent of $w^\alpha, \bar w^{\dot \alpha}$. In local coordinates $(w,1)\sim(w^0,w^1)$, combined with the inhomogeneous part $\bar\partial \chi_{\text{ST}} = -2 v \bar \partial^2_0 f\bar e^0$, the leading order term in the large $r$ expansion of this vector field acting on an on-shell state is
\begin{equation}
    \delta_{\chi_{ST}}
    h = h + \bar \partial \chi_{ST} + f(w,\bar w)\frac{\partial h}{\partial u} + \mathcal{O}\left(\frac{h}{r^{1/2}}\right)\,,
\end{equation}
the homogenous part of the transformation is indeed the action of the supertranslations with the supertranslation profile $f(w,\bar w)$. $\chi_{\text{ST}}$ is defined in \eqref{eq: gauge parameters for ST, SR}. The superrotation calculation is somewhat more involved but essentially follows from the same procedure using the profile $\chi_{\text{SR}}$ defined in \eqref{eq: gauge parameters for ST, SR}. The final result is
\begin{multline}    
    \delta_{\chi_{SR}} h = h+ \bar \partial \chi_{SR}+ \left(Y \bar \partial_0 h +\bar \partial_0 Y(1 + \frac{1}{2}u \partial_u) h\right) 
    \\ + \mathcal{O}\left(\frac{h}{r^{1/2}}\right)\,,
\end{multline}
in which $\bar \partial \chi_{SR} = -\frac{1}{2}v^2 \bar \partial_0^3 Y \bar e^0$\footnote{It is interesting to note that the inhomogenous part of the supertranslation and the superrotation in fact agrees with the known results if we take the Newman prescription $v = u \iff \lambda^\alpha \rightarrow w^\alpha$ and
\begin{align}
    \delta_{\text{inhom}} \bar\sigma \la\lambda\hat\lambda\ra \D\hat\lambda &= \la\lambda\hat\lambda\ra \partial_{u_{Newman}} \bar \partial \chi =  \partial_u \bar \partial \chi
    \\
    \delta_{\text{inhom} ST} \bar \sigma &= -2 \bar \partial_0^2 f
    \\
    \delta_{\text{inhom} SR} \bar \sigma &= -u \bar \partial_0^3 Y
\end{align}
although the interpretation of this computation in our context is less clear.}. The coefficient in the $\bar \partial_0 Y h$ term is $1$ rather than $-\frac{1}{2}$ in, for example, \cite{Barnich:2010ojg,Pasterski_2019}. This is likely because we are working with the $(0,1)$-form representatives and not the metric components, which are related by multiple derivatives and an integral transform. The fact one recovers asymptotic shear in the large $r$ limit was also hinted in the derivations of \cite{Donnay:2024qwq}. Similar computations have been done in \cite{Kmec:2024nmu} where both generators and their higher order descendants were written down with asymptotic twistor space.

\section{Stress tensor}\label{sec:stress_tensor}
The form of the holomorphic stress tensor can be read off from the PCM-like action \eqref{2d_action} as the current that couples to the following infinitesimal complex structure perturbation on the Riemann sphere,
\begin{equation}
    \bar\partial\to\bar\partial+ b\,\partial\,,
\end{equation}
where the Beltrami differential $b\in\Omega^{0,1}(\mathbb{CP}^1,T_{\mathbb{CP}^1})$ is an infinitesimal deformation of the complex structure. The holomorphic stress tensor sourced by the Beltrami is given by the object coupling to $b$:
\begin{multline}
    T(\lambda,\bar\lambda)= \Phi_{\boldsymbol{a}}\partial\alpha^{\boldsymbol{a}}+\Lambda_{\boldsymbol{a}}\partial \beta^{\boldsymbol{a}}+ \partial\zeta_{\ba}\partial\zeta^{\ba} 
    \\ +  (\Delta-s)\partial\left(\Phi_{\boldsymbol{a}}\alpha^{\boldsymbol{a}}+\Lambda_{\boldsymbol{a}}\beta^{\boldsymbol{a}}\right) \,.
\end{multline}
The total derivative term comes from the fact that the fields were sections of $\mathcal{O}(*)$. The $\Delta$ and $s$ in the coefficient of the total derivative term are the mode numbers inside $\ba =(\Delta,s,k)$. We remind the reader that indices are raised and lowered with $\kappa^{\boldsymbol{ab}}$ as defined in \eqref{eq: algebra constants}, so upstairs and downstairs $\ba$ differ by a sign. The $TT$ self-OPE has the expected form with central charge $c=2\kappa^2\propto \text{vol}(M)$, which diverges. It is easy to check that this does generate the stress tensor OPE, written in local coordinates as:
\begin{equation}
\begin{aligned}
    &T(\lambda_1,\bar\lambda_1) J^{\boldsymbol{a}}(\lambda_2,\bar\lambda_2)
    \\\sim &\left(\frac{h}{(z_1-z_2)^2}-\frac{\partial_{z_2}}{z_1-z_2}\right) J^{\boldsymbol{a}}(\lambda_2,\bar\lambda_2)\,.
\end{aligned}
\end{equation}
But this is only a chiral half of the 2d stress tensor, which is not sourced by a bulk graviton but rather a Beltrami differential. This is the consequence of working with a chiral theory that encodes the bulk gravitons as $(0,1)$-forms. There is no obvious canonical way to arrive at an object that can encode an anti-holomorphic Celestial CFT stress tensor.

To start with, we consider the cubic coupling term in our action on the momentum space celestial sphere \eqref{2d_action}:
\begin{equation}
    \int_{\mathbb{P}^1} \D\lambda \, h^{\ba}\left(f^{\bb}_{\ba\bc}\, \Phi_{\bb} \alpha^{\bc} + ... \right)(\lambda)\,.
\end{equation}
Here the omitted part $...$ includes the mass term (non-bilinear) terms in equation \eqref{eq: currents}. In the spirit of usual bulk-boundary correspondence, bulk gravitons couple to stress tensor, although we are only on the momentum space celestial sphere, we see that the boundary stress tensor sourced by a graviton should in principle come from the bracketed part \footnote{Note that $J_{\Delta,s}$ generates the Witt algebra, it is perhaps not surprising that the graviton modes that couple to it should induce a stress tensor after summing over all the modes.}. For notational convenience, we undo the mode decomposition 
\begin{equation}        f^{\bb}_{\ba\bc}\Phi_{\bb}\alpha^{\bc}\leftrightarrow \left\{\Phi,\alpha \right\} (x, \lambda)\,,
\end{equation}
where the bracket is defined in \eqref{PB_x} and the fields without index depend on spacetime. For $\Phi, \Lambda$, the distributional edge modes, we have
\begin{equation}
    \Phi_{\Delta,s,k} \to \Phi = \int_{\Delta}\d\Delta\,\sum_{s,k}\,B_{\Delta,s}\,{\delta^{k-1}(v)\delta(v')}\,\Phi_{\Delta,s,k}\,.
\end{equation}
Undoing the mode decomposition on $h_{\Delta,s,k}, \alpha_{\Delta,s,k} ...$ give, for instance:
\begin{equation}
    h = \int \d \Delta \sum_{s,k} B_{\Delta,s} v^k h_{\Delta,s,k} \,,
\end{equation}
in which the lack of $v'$ dependence comes from the fact that $v'$ can be ignored on the support of the $\delta$-functions in the definition of $\Lambda$.

In this index free notation, the cubic coupling term in the 2d action takes the form:
\begin{equation}
    \int_{\mathbb{P}^1\times\mathbb{M}} \D\lambda\wedge \d^4x\, h\, \underbrace{\, \left(\left\{\Phi,\alpha\right\})(\lambda,x) + ...\right) }_{\text{boundary current}}\,,
\end{equation}
where the term in under brace indicates the boundary source for a bulk graviton mode. From the free-field OPEs of the modes, it can be worked out \cite{Seet:2024vmh} that the index-free notation version of \eqref{JJOPE} is
\begin{align}\label{eq: Poisson algebra of currents}
    &J_f(\lambda) := \int \d^4 x f(x,\lambda)\left(\{\alpha,\Phi\}+\{\beta,\Lambda\} + ... \right) \nonumber
    \\
    &\frac{1}{2 \pi \im}\oint_{|z_1-z_2|=\epsilon} \d z_1 J_{f_1}(z_1) J_{f_2}(z_2) = J_{\{f_1,f_2\}}(z_2)
\end{align}
where we have given the last line in local coordinates, and the simple pole between the currents takes the Poisson brackets between the $f_1,f_2$. 

In order for the stress tensor to emerge, we consider a global Kerr-Schild type deformation \cite{Kerr2009}, which takes the form of the following linearized metric perturbation in Minkowski space. The most generic form of such perturbations in 4d amounts to choosing a constant complex null vector $K^{\mu}$ such that:
\begin{equation}
    \delta g^{\mu\nu}= \varepsilon \partial^{(\mu}\left(K^{\nu)}(x\cdot K)\right)= \varepsilon K^{(\mu}K^{\nu)} \,,
\end{equation}
induced by a coordinate transformation $\delta x^{\mu}=\varepsilon K^{\mu}(x\cdot K)$ where $K^{\mu}=\hat{k}^{\alpha}\bar k^{\dal}$ such that $K\cdot T=0$. It is important to note that this is not a Lorentz transformation since we are not anti-symmetrizing over $\mu$ and $\nu$ (which would be $0$), but rather a constant complex Kerr-Schild deformation that preserves the null condition on $\partial_u$ (the $uu$-component of the metric perturbation vanishes). Via the integral transform, it can be found that the counter part of this Minkowski space diffeomorphism takes the simple form
\begin{equation}
    \int_{\mathbb{CP}^1}\D\lambda\,\frac{\hat{\lambda}^{\alpha}\hat{\lambda}^{\beta}\hat{\lambda}^{\gamma}\hat{\lambda}^{\delta}}{\la\lambda\hat{\lambda}\ra^4}\,\frac{\partial^2}{\partial x^{\gamma\dal}\partial x^{\delta\dbeta}} \underbrace{\left(\frac{\la\lambda|x|\bar\lambda]^2}{[\bar k\bar\lambda]^4}\,\D\bar\lambda \right)}_{\text{$\mathbb{P}^1$ perturbation} }\,,
\end{equation}
where the underbraced term gives a cohomological gauge transformation on the momentum space celestial sphere:
\begin{equation}
    \delta h = \frac{\la\lambda|x|\bar\lambda]^2}{[\bar k\bar\lambda]^4}\,\D\bar\lambda\,. 
\end{equation}
This induces the definition of the boundary stress tensor:
\begin{equation}
\begin{aligned}
    &\bar T=\int_{\mathcal{M}^\cO\times \mathbb{CP}^1}\D\lambda\d^4 x\,\frac{\delta S_{2d}}{\delta h}\,\frac{\la\lambda|x|\bar\lambda]^2}{[\bar k\bar\lambda]^4}\,\D\bar\lambda\\
    =&\int_{\mathcal{M}^\cO\times \mathbb{CP}^1}\D\lambda\d^4 x\,\left(\{\alpha,\Phi\}+\{\beta,\Lambda\} + ... \right)\,\frac{\la\lambda|x|\bar\lambda]^2}{[\bar k\bar\lambda]^4}\,\D\bar\lambda\,.
\end{aligned}
\end{equation}
In particular, the action of the candidate stress tensor on a self-dual or an anti-self-dual hard graviton:
\begin{multline}
    \int_{\mathbb{P}^1_{\lambda_1}}\int_{\mathcal{M}}\d^4x_1\frac{\la\lambda_1|x_1|\bar\lambda_1]^2}{[\bar k\bar\lambda_1]^4}\,(\{\alpha,\Phi\}+\{\beta,\Lambda\}+...)(\lambda_1,x_1)\,\\ \times \mathcal{J}_{k}(\lambda_2,\bar\lambda_2)
    \sim \\ \int_{\mathbb{P}^1_{\lambda_1}}\frac{\D\lambda_1\D\bar\lambda_1}{[\bar k\bar\lambda_1]^4}\underbrace{\frac{[\bar\lambda_1\bar\lambda_2]^2\bar\partial_{\bar\lambda_2}-\Delta[\bar\lambda_1 \bar\lambda_2]}{\la\lambda_1\lambda_2\ra}}_{C(\lambda_1,\bar\lambda_1)}\, \mathcal{J}_{k}(\lambda_2,\bar\lambda_2)\,,
\end{multline}
where the (in this case anti-self-dual) hard graviton can be written as bilinears of the free fields we have introduced (see Appendix \ref{appendix: hard particles}):
\begin{equation}
    \mathcal{\tilde J}_{k}=\int_{\mathcal{M}}\d^4x
    \frac{1}{\la\lambda|x|\bar\lambda]^{k}}\,\left\{\Lambda,\alpha\right\}\,.
\end{equation}
We have used the holomorphic OPE between the free fields to generate the simple pole in $\la\lambda_1\lambda_2\ra$. 
We recognize the remaining integral as taking a shadow transform as defined in \cite{Simmons-Duffin:2012juh,Cheung:2016iub,Banerjee:2022wht}. And the resulting OPE coefficients gives the desired anti-holomorphic stress tensor.
\begin{equation}\label{eq: TbarOPE}
    \bar T(k^{\alpha},\bar k^{\dal}) \mathcal{J}_{k}(\lambda_2)\sim \left(\frac{k}{[\bar k\bar\lambda_2]^2}-\frac{\bar\partial_{\bar\lambda_2}}{[\bar k\bar\lambda_2]}\right) \mathcal{J}_{k}(\lambda_2)\,.
\end{equation}
We check this explicitly in the appendix. The $\bar T\bar T$ self-OPE is slightly subtle to compute, as done in section 4.1 in \cite{Fotopoulos_2020}. Due to the presence of the non-local integral in the definition of $\bar T$, one is forced to make the two variables integrated over coincide to obtain the $\bar T\bar T$ OPE, which is a rather subtle procedure. Given the action of the antiholomorphic stress tensor on a current \eqref{eq: TbarOPE}, we simply open up the definition of the second $\bar T$ as an integral transform of the subleading soft graviton and act the first $\bar T$ on it. The terms self-organize to give the expected stress tensor OPE without a quartic pole:
\begin{equation}
    \bar T(k_1^{\alpha},\bar k_1^{\dal})\,\bar T(k_2^{\alpha},\bar k_2^{\dal})\sim \frac{2\bar T(k_2^{\alpha},\bar k_2^{\dal})}{[\bar k_1\bar k_2]^2}-\frac{\bar\partial_{\bar k_2} \bar T(k_2^{\alpha},\bar k_2^{\dal})}{[\bar k_1\bar k_2]}  \,.
\end{equation}
Indicating the central charge $\bar c=0$ from the antiholomorphic sector of the 2d theory.

\section{Dressing}\label{sec:dressing}
We have seen in section \ref{sec:ST/SR} that it is straightforward to construct charges associated with supertranslation and superrotation using arbitrary profile functions of $\lambda,\bar\lambda$ integrated against particular linear combinations of the currents $J^{\Delta,s,k}$.

One of the currents used to build $Y_{\text{SR}}$ given in \eqref{eq: supertranslation and superrotation currents} is $J^{1,1,-1}$, which has the distinguishing property that it acts diagonally on all other operators built from $\alpha,\Phi,\beta,\Lambda$ (up to potential double pole terms if the mode numbers of $\mathcal{O}$ coincide with $(1,1,-1)$)
\begin{equation}
    J^{1,1,-1}(z_1) \mathcal{O}^{\Delta,s,k}(z_2) \sim \frac{-\frac{1}{2}(\frac{1}{2}(\Delta+s) + k)\mathcal{O}^{\Delta,s,k}(z_2)}{z_1-z_2} \,.
\end{equation}
Another way of seeing this is that in the index-free form of the currents, $J^{1,1,-1}$ came with the $\mathcal{M}$ dependent mode function $qv$. Under the Poisson bracket as in \eqref{eq: Poisson algebra of currents}, this generates the Hamiltonian vector field $1/2(q\partial_q-v\partial_v)$, which measures the combination of mode numbers $\frac{1}{2}(\Delta+s) + k$.

$J^{1,1,-1}$ is roughly speaking a generator of the Cartan subalgebra of this current algebra. The linearised gauge variation of the graviton $\delta h$ that generates $J^{1,1,-1}$ has $x$-dependence of $qv$, corresponding on spacetime via \eqref{eq: vector field}, \eqref{Sparling} to a large ($\mathcal{O}(r^0)$, does not fall off at large $x$) diffeomorphism.

Classically, charges generated by $J^{1,1,-1}$ obey an Abelian fusion rule, allowing us to construct a family of $U(1)$ charges
\begin{align}
    Q_\phi:=\text{Exp}\left(\im \int \D\lambda \, (\phi \bar\partial J^{1,1,-1})\right)
    \\
    Q_\phi \mathcal{O}^{\Delta,s,k}(z) \sim e^{-\im \frac{1}{2}(\frac{\Delta+s}{2}+k)\phi}\mathcal{O}^{\Delta,s,k}(z)\,.
\end{align}
In particular, our 2d action \eqref{2d_action} is not invariant under the action of these charges. This is because these charges act only on the fields $\Phi,\Lambda,\alpha,\beta$, and the compensating gauge transformation of the bulk graviton mode $h^{\ba}$
\begin{equation}
    h^{\Delta,s,k} \rightarrow h^{\Delta,s,k}e^{-\im \frac{1}{2}(\frac{\Delta+s}{2}+k)\phi}
\end{equation}
is not generated by this charge. Each individual choice of the phase $\phi$ associated with the $U(1)$ transformation picks out a particular leading $1/r$ coefficient in a asymptotic expansion of the bulk metric. This breaks the infinitely degenerate asymptotic vacuum for linearised gravity, which in our action, is manifested as the requirement for the compensating gauge transformation of the bulk graviton mode.

This naturally motivates us to consider asymptotic gravitons which are sums over all such possible vacuum choices. This is precisely the idea of dressing asymptotic states with clouds of linearised long range interactions, i.e. soft gravitons \cite{Donnelly:2015hta,Strominger:2017zoo}. For example, consider the following dressed state $\cO^{\ba}(z,\bar z)$, which corresponds to some gravitationally coupled particle in the bulk:
\begin{equation}
 \cO^{\ba}(z,\bar z) \,\text{Exp}\left(\im Q\int_{\mathbb{CP}^1}\,\frac{\D\lambda'\la \iota z\ra}{\la z\lambda'\ra\la\iota\lambda'\ra}\,h^{0}\right)\,,
\end{equation}
where $Q = \frac{1}{2}(\frac{\Delta+s}{2}+k)$ is the charge of $\cO^{\ba}(z,\bar z)$ under $J^{1,1,-1}$, and the exponentiated integrand is the soft graviton mode $h^0:= h^{1,1,-1}$ (where we recall that $h^{1,1,-1}qv \in h \in \Omega^{0,1}(\mathbb{CP}^1,\mathcal{O}(2))$. Since $q \in \mathcal{O}(2)$, $h^{1,1,-1} \in \Omega^{0,1}(\mathbb{CP}^1,\mathcal{O})$) with a reference spinor $\iota^\alpha$ amounting to the choice of the endpoint of the dressing line on the celestial sphere. The linear part of the diffeomorphism of $h$ is
\begin{equation}
    h^0 \to h^0+\bar\partial a\,,
\end{equation}
where $a\in\Omega^0(\mathbb{CP}^1,\mathcal{O})$ is not free but is constrained to have at least a simple zero at $\iota^\alpha$, the endpoint of the dressing line.

Via the non-local integral transform \eqref{Sparling}, we see that this is the linearized diffeomorphism:
\begin{equation}
    g_{\mu\nu} \to g_{\mu\nu} + 2\partial_{(\mu}\zeta_{\nu)}\,.
\end{equation}
Under the action of this $U(1)$ (closely associated to superrotations) which indicates linearized gravity vacuum transition, the bare gravitationally coupled operator $\cO^{\ba}(z,\bar z)$ transforms as 
\begin{equation}
    \cO^{\ba}(z,\bar z) \to \cO^{\ba}(z,\bar z) \e^{-\im Q a}\,,
\end{equation}
Hence the dressed combination is invariant under the $U(1)$ transformation:
\begin{equation}
\begin{aligned}
    &\cO_k(z,\bar z)\e^{\im Q\bar\partial^{-1}h^{0}} \rightarrow \\
    & \left(\cO_k(z,\bar z)\,\e^{-\im Q a}\right)\left(\e^{\im Q\bar\partial^{-1}h^{0}}\e^{\im Q a}\right)\,.
\end{aligned}
\end{equation}

\section{Discussion}
We have identified the independent gravitational edge mode degrees of freedom living on the 2d celestial sphere in an explicit Lagrangian form, from which one can explicitly construct soft dynamics of 4d gravity. This short manuscript serves as a further elucidation of the previous paper where we derived the action functional \cite{Bu:2024cql}. It leaves several interesting avenues for further investigations. 

The fact that there are infinitely many fields in the theory is obviously indicating the theory is actually higher dimensional, we force it to be in 2d for the convenience of using 2d CFT techniques to do calculations. To recover the 3d theory, one must restore the mode functions that carried the spatial dependence on the extra dimension. It would be interesting to see if choices of mode numbers and mode functions to restore will allow the action functional to be extended to 3d manifolds, either $\mathbb{R}\times\mathbb{CP}^1$ or $S^3$. 
The latter requires us to introduce an additional parameter $l$, namely the radius of the circle $S^1$ in the Hopf fibration, taking $l\to\infty$ allows for a closer study of the transition from 3d boundary of $AdS_4$ to null boundary of 4d asymptotically flat spacetimes. Admittedly a subtle limit, but nevertheless might help relating the edge modes of Einstein gravity in the two spacetimes.

Given the 2d action for vacuum Minkowski space, it would be interesting to compute its partition function, one expects this to count the number of allowed leading metric component in the fall off, which should diverge. However, one could also run the derivation procedure and compute the partition function for other spacetimes such as self-dual Taub-NUT \cite{Guevara:2023wlr,Bogna:2024gnt}, the quotient of the two partitions would be finite, acting as a vacuum subtraction procedure for the curved spacetime. The quotient computes the radiation entropy for the curved space.

\acknowledgments
It is a pleasure to thank Temple He, Joon-Hwi Kim, Jacob Mcnamara, Natalie Paquette and David Skinner for interesting discussions. We also thank Atul Sharma for commenting on the draft. WB is supported by the Royal Society Studentship and the Simons Collaboration on celestial holography. SS is supported by the Trinity Internal Graduate Studentship and the Simons Collaboration on celestial holography. The work of SS has been supported in part by STFC HEP Theory Consolidated grant ST/T000694/1.

\newpage
\newpage

\bibliographystyle{apsrev4-1}
\bibliography{CCFT}

\newpage

\appendix
\section{The twistorial derivation of the PCM-like theory}\label{appendix: twistor derivation}
This family of PCM-like actions was systematically derived from the twistor actions for self-dual Yang-Mills and for self-dual gravity by introducing new fields $\Phi, \Lambda$ associated with edge mode degrees of freedom. A full derivation is available in \cite{Bu:2023cef,Bu:2024cql}, but the necessity of including new fields in the twistor action when studying celestial holography is made clear by a careful study of the twistor space scattering states. For example, (taking the self-dual gravity case for definiteness) the twistor space scattering states that correspond to conformal primary states have wavefunctions of the form
\begin{equation}
    h_{\text{on-shell}} \propto \bar \delta(\la \lambda \lambda_i \ra)\left(\frac{\la \lambda \hat \lambda\ra}{[\mu \bar \lambda]}\right)^{k}, \quad Re(k) > 0
\end{equation}
and do not obey the linearised equation of motion $\bar \partial h_{\text{on-shell}} = 0$. Rather, they obey a sourced equation of motion
\begin{equation}
    \bar \partial h_{\text{on-shell}} \propto \bar \delta([\mu \bar \lambda]),
\end{equation}
leading to an inconsistency in identifying the scattering states as solutions to the linearised equations of motion. In \cite{Bu:2023cef, Bu:2024cql} this is remedied by replacing every appearance of $h$ in the twistor self-dual gravity action with $h + \bar \delta^{(k)}([\mu \bar \lambda]/\la \lambda \hat \lambda \ra)\Lambda_k$, and likewise for $\tilde h$.
\begin{align}
S[h, \tilde h]:= \int \D^3 Z \tilde h \left(\bar \partial h + \frac{1}{2}\{h,h\}\right) +S_{\text{gauge fixing}}
\nonumber\\
\rightarrow S\left[h + \bar \delta^{(k)}\left(\frac{[\mu \bar \lambda]}{\la \lambda \hat \lambda \ra}\right)\Lambda_k,\tilde h + \bar \delta^{(k)}\left(\frac{[\mu \bar \lambda]}{\la \lambda \hat \lambda \ra}\right)\Phi_k \right]\,.
\end{align} 
This corrects the linearised equations of motion for $h$ to
\begin{equation}
    \bar \partial\left(h + \bar \delta^{(k)}([\mu \bar \lambda]/\la \lambda \hat \lambda \ra)\Lambda_k\right) = 0\,.
\end{equation}From this point of view, $\Lambda_k, \Phi_k$ are to be interpreted as sources or real-codimension-2 edge modes.

The theory of $\Phi_k, \Lambda_k, h, \tilde h$ supported on the real-codimension-2 locus $\{[\mu \bar \lambda]=0\} \subset \{\mathbb{PT}\setminus\{\mu^{\dot \alpha}=0\}\}$ is an alternative presentation of the PCM-like action \eqref{2d_action}. To get to the form given in \eqref{2d_action}, we perform a mode decomposition in the remaining $\mu^{\dot \alpha}$ dependence. In local coordinates, it can be shown that the remaining $\mu^{\dot \alpha}$ dependence of the quantum fields is on one parameter taking values in $\mathbb{C}^* \cong \mathbb{R}^+ \times U(1)$, and therefore a mode decomposition in this variable contributes a continuous mode number $\Delta \in 1 + \im \mathbb{R}$ (by a normalizability argument, taking the Fourier transform in the log of the $\mathbb{R}^+$ coordinate) and a discrete mode number $s \in \mathbb{Z}$ (taking the Fourier series in the $U(1)$ coordinate). Altogether, we recover a theory on $\mathbb{CP}^1_{\lambda}$ with fields labeled by 3 mode numbers $(k, \Delta, s)$
\begin{equation}
   k, \{\{[\mu\bar \lambda]=0\} \subset \mathbb{PT}\setminus\{\mu^{\dot \alpha}=0\}\} \rightarrow (k, \Delta, s), \mathbb{CP}^1_{\lambda}\,.
\end{equation}
On the left hand side we have the original coordinates and mode numbers our fields $h, \tilde h, \Phi_k, \Lambda_k$ depend on, while on the right hand side we have that the fields depend on the coordinates $\lambda^{\alpha}$ of the $\mathbb{CP}^1$, as well as mode numbers $(k,\Delta,s)$, which we collectively denote as $\boldsymbol{a} = (k, \Delta, s) \in (\mathbb{Z}, 1+ \im \mathbb{R}, \mathbb{Z}_+)$. One important thing to note is that as per the definition of $\Phi_k, \Lambda_k$, the $k$ mode number on them is strictly positive (labeling the strength of the pole in $[\mu \bar \lambda]$), but the $k$ label on all other fields is permitted to be arbitrary. This is because there are cubic terms in the action, and the orthogonality rule for $k$ arises from the canonical pairing of functions and distributions
\begin{equation}
    \int \frac{\d v}{v} v^{k_1} v^{k_2} \bar\delta^{(k_3-1)}(v)
    = (-1)^{k_3-1} (k_3-1)! \delta_{k_1+k_2,k_3}\,,
\end{equation}
where the $v=[\mu \bar \lambda]/\la \lambda \hat \lambda \ra$ integral above comes from integrating out the $v$ dependence of 3 quantum fields appearing in a cubic interaction term. We see that requiring that $k_3$ is positive only constrains the sum of $k_1+k_2$, and not each individually. Tracing through the integrals and mode decompositions carefully to go from the sub-theory on $[\mu \bar \lambda]=0$ to the PCM-like action, we recover the following definitions
for the metric and structure constants
\begin{align}
    &\kappa^{\boldsymbol{a}_1\boldsymbol{a}_2} = \kappa_{\boldsymbol{a}_1\boldsymbol{a}_2}:= \delta(\Delta_1+\Delta_2)\delta_{s_1+s_2,0} \delta_{k_1+k_2,0} 
    \\
    & f^{\boldsymbol{a}_1\boldsymbol{a}_2\boldsymbol{a}_3} := \delta(\Delta_1+\Delta_2+\Delta_3-1)\delta_{(s_1+s_2+s_3+1),0}\times \nonumber
    \\ &\delta_{(k_1+k_2+k_3-1),0}\left(k_2\left(\frac{s_3-\Delta_3}{2}\right) - k_3\left(\frac{s_2-\Delta_2}{2}\right)\right)
    \\ & c(\boldsymbol{a}):= \frac{\Delta-s}{2}\,.
\end{align}
The form components of $h$ and $\tilde h$ that appear in the action are in modes of the $(0,1)$-forms, one simply expands the fields in basis $B_{\Delta,s}$ and $1/u^k$ as mentioned in the main text, this gives a composite number $\ba=(\Delta,s,k)$ labeling the mode. 


\section{Detail of the stress tensor computation}
In local coordinates, the Mellin transform of a Fourier transform of the momentum space celestial sphere stress tensor OPE coefficient looks like:
\begin{equation}
    \int_{\mathbb{C}_{w}}\frac{\d w\d\bar w}{(w-\bar z)^4}\frac{(w-\bar z_2)^2\bar\partial_{\bar z_2}-\Delta(w-\bar z_2)}{(\bar w-z_2)}\,,
\end{equation}
where we have identified $\bar\lambda_1^{\dal}=\bar k_1^{\dal}$ and $\bar\lambda_2^{\dal}=\bar k_2^{\dal}$. We have picked $\bar k_1^{\dal}=\binom{1}{w}$, $\bar k_2^{\dal}=\bar\lambda_2=\binom{1}{\bar z_2}$, $\bar\lambda_1^{\dal}=\binom{1}{\bar z}$ and their corresponding complex conjugates. The integral can be done by shifting the argument by $w\to w-\bar z_2$:
\begin{equation}
    \int_{\mathbb{C}_{w}}\frac{\d w\d\bar w}{(w-(\bar z-\bar z_2))^4}\frac{w^2\bar\partial_{\bar z_2}-\Delta w}{\bar w}\,.
\end{equation}
Parametrizing $w=r \e^{\im\theta}$ and $\bar w=r\e^{-\im\theta}$ allows us to do the radial integral:
\begin{equation}
\begin{aligned}
    &\int_{0}^{2\pi}\d\theta \int_{0}^\infty \d r \,\frac{r^2\bar\partial_{\bar z_2}\e^{3\im\theta}-\Delta\,r\,\e^{2\im\theta}}{(r\e^{\im\theta}-(\bar z-\bar z_2))^4} \\
    &= \frac{2\pi}{3}\left(-\frac{\bar\partial_{\bar z_2}}{(\bar z-\bar z_2)}+\frac{\Delta}{2(\bar z-\bar z_2)^2} \right)\,.
\end{aligned}
\end{equation}
\section{Hard particles as current insertions}\label{appendix: hard particles}
On twistor space, the on-shell wavefunctions that correspond to the conformal primary basis can be derived by taking the Mellin transform of the standard plane wave representatives. Define $h_{\text{on-shell}(\Delta, \lambda_p)}=$:
\begin{align}
    \int \d \omega \omega^{\Delta -1 } \left(\bar \delta_{2,-4}(\lambda^{\alpha}, \sqrt{\omega}\lambda_p^{\alpha}) \exp \left(\im \sqrt{\omega}\frac{[\mu \bar \lambda]}{\la \lambda \hat \lambda \ra}\right)\right)\nonumber
    \\
    =\frac{\Gamma(\Delta-3)\,\bar \delta(\la\lambda \lambda_p\ra)\la\lambda\iota\ra^3}{[\mu \bar \lambda]^{\Delta-2}\la\lambda_p\iota\ra^3} \la \lambda \hat \lambda \ra^{\Delta-2}\,.
\end{align}
In the full twistor theory, computing an amplitude with an external hard particle $h_{\text{on-shell}(\Delta, \lambda_p)}$ is equivalent to computing a correlator with the insertion of the current that this wavefunction sources
\begin{equation}
    \mathcal{J}_{\Delta}(\lambda_{p}) e^{S_{int}} = \left(h_{\text{on-shell}(\Delta, \lambda_p)},\frac{\delta S_{int}}{\delta h}\right) e^{S_{int}}\,.
\end{equation}
$S_{int}$ are the interaction vertices of the action, which in this case are the cubic terms. This prescription is a minor modification of the Berends-Giele current method of computing Feynman diagrams through recursion and can be shown to be equivalent to the standard LSZ prescription diagram-by-diagram in Feynmann diagrammatics. It is however more suited to our needs because the amplitude calculation manifestly becomes the computation of a current correlator. OPE rules are supplied by Wick contractions in the familiar way.
    \section{Localisation and identification of celestial sphere with Bondi sphere}\label{appendix: large r delta functions}
For $b \in \mathbb{Z}^+, Re(a)>b/2$, the leading order term in the large $r$ limit is
\begin{equation}
    \frac{r^{b/2}}{(u+ r \frac{\rho^2}{1+\rho^2})^a} \propto u^{b/2-a}\left(\frac{\partial}{\partial \rho^{b-1}}\delta(\rho)\right)\frac{\Gamma(a-b/2)}{\Gamma(a)\Gamma{(b)}}\,,
\end{equation}
with a proportionality constant independent of $a,b$ of order 1. The proof is as follows. Consider a new variable $\rho' = \sqrt{r/u}\,\rho$, with $b \in \mathbb{R}^+$:
\begin{multline}
    \int_{0}^{\infty} \d \rho \left(\frac{r^{b/2}\rho^{b-1}}{(u+ r \frac{\rho^2}{1+\rho^2})^a}\right) f(\rho) 
    \\= \int_{0}^{\infty} \d \rho' \left(\frac{(\rho')^{b-1}}{(1+ \frac{(\rho')^2}{1+\frac{u}{r}(\rho')^2})^a}\right) f(\rho'\sqrt{\frac{u}{r}})\,.
\end{multline}
In the large $r$ limit with fixed $u$, we have
\begin{equation}
    = f(0)\int^{\infty}_{0}\d \rho \frac{\rho^{b-1}}{(1+\rho^2)^a} + \mathcal{O}(1/r^{1/2})\,.
\end{equation}
Where we see that the use of $r\rho^2/(1+\rho^2)$ in the denominator instead of $r\rho^2$ was irrelevant in the large $r$ limit. For convergence of the final integral, we require the constraint
\begin{equation}
    Re(a) \geq \frac{b}{2} \,,
\end{equation}
In the integrals that we will consider, the measure for the $\rho$ integral comes from a $\int_{\mathbb{C}}\d^2 z = \int_{\mathbb{R}^2} \rho \d \rho \d \theta$. The case of $b=2$ is therefore of special interest. One useful result is
\begin{equation}
    \int_0^\infty \rho \d \rho \int_0^{2 \pi} \d \theta \frac{f(\rho e^{\im \theta}, \rho e^{-\im \theta})}{(u+ r \frac{\rho^2}{1+\rho^2})^a}  = \pi \frac{1-a}{r u^{a-1}} f(0,0) \,.
\end{equation}
 
\section{Integral transform for vector fields}\label{appendix: Penrose transform for vectors}
The vector fields that generate infinitesimal diffeomorphisms in flat space can be constructed from the symmetry parameter that generates linearised cohomology transformations in the Mason-Wolf twistor action \eqref{twistor sdg}. To construct the dictionary, consider the (antipodal gauge choice for the) Penrose transform to the metric perturbation around flat space
\begin{equation}
    \delta g_{\alpha \dot \alpha \beta\dot \beta} = \int \D \lambda \frac{\hat \lambda_{\alpha} \hat \lambda_\beta}{\la \lambda \hat \lambda \ra^2} \mathcal{L}_{\frac{\partial}{\partial \mu^{\dot \alpha}}}\mathcal{L}_{\frac{\partial}{\partial \mu^{\dot \beta}}} \bar \partial \chi\,,
\end{equation}
where $\mathcal{L}_{\frac{\partial}{\partial \mu^{\dot \alpha}}}$ denotes the Lie derivative along $\mu^{\dal}=x^{\alpha\dal}\lambda_{\alpha}$.
For the special case in which $\bar \partial \chi \propto \D \hat \lambda$, i.e. the symmetry parameter $\chi$ is holomorphic in $\mu^{\dot \alpha}$, we have the simple relation that
\begin{equation}
    \bar \partial \chi = \bar \partial_{\lambda} \chi\,.
\end{equation}
Substituting this expression into the integral transform and integrating by parts, we have
\begin{equation}
    \delta g_{\alpha \dot \alpha \beta\dot \beta} = 2 \int \frac{\D \lambda \D \hat \lambda}{\la \lambda \hat \lambda \ra^2} \frac{\lambda_{(\alpha} \hat \lambda_{\beta)}}{\la \lambda \hat \lambda \ra^2} \mathcal{L}_{\frac{\partial}{\partial \mu^{\dot \alpha}}}\mathcal{L}_{\frac{\partial}{\partial \mu^{\dot \beta}}} \chi\,.
\end{equation}
Since $\lambda_\alpha \partial_{\dot \alpha}$ agrees with $\partial_{\alpha \dot \alpha}$ when acting on $\chi$, we can instead write this expression as
\begin{equation}
    \delta g_{\alpha \dot \alpha \beta\dot \beta} = 2 \partial_{(\dot \alpha| (\alpha} \int \frac{\D \lambda \D \hat \lambda}{\la \lambda \hat \lambda \ra^2} \frac{\hat \lambda_{\beta)}}{\la \lambda \hat \lambda \ra^2} {\frac{\partial}{\partial \mu^{\dot \beta)}}} \chi(\mu, \lambda, \bar \lambda)\,,
\end{equation}
where we have replaced the Lie derivative with its action on scalars, the ordinary derivative. The final integral is recognizably the component of the vector field that generates the infinitesimal diffeomorphism.
\begin{equation}
    V_{\beta \dot \beta} = \int \frac{\D \lambda \D \hat \lambda}{\la \lambda \hat \lambda \ra^2} \frac{\hat \lambda_{\beta}}{\la \lambda \hat \lambda \ra^2} {\frac{\partial}{\partial \mu^{\dot \beta}}} \chi(\mu, \lambda, \bar \lambda)\,.
\end{equation}
and the full vector field therefore takes the form
\begin{equation}
    V = \int \frac{\D \lambda \D \hat \lambda}{\la \lambda \hat \lambda \ra^2} \{\chi(\mu, \lambda, \bar \lambda), \,\cdot \}\,.
\end{equation}

\end{document}